\documentclass[sigconf, nonacm]{acmart}
\usepackage[english]{babel}
\usepackage{array}

\renewcommand\footnotetextcopyrightpermission[1]{} % removes footnote with conference info
\setcopyright{none}
\usepackage{booktabs} % For formal tables
\usepackage{tablefootnote}
\usepackage{bbding}
\usepackage{quoting}
\quotingsetup{vskip=1pt}
\quotingsetup{leftmargin=10pt}
\quotingsetup{rightmargin=10pt}

\begin{document}

\title{Design Considerations for Low Power Internet Protocols}

%\author{Paper \# 65 (12 pages)} %Limit is 12 pages including citations
\author{Hudson Ayers}
\affiliation{%
  \institution{Stanford University}
}
\email{hayers@stanford.edu}

\author{Paul Thomas Crews}
\affiliation{%
  \institution{Stanford University}
}
\email{ptcrews@stanford.edu}

\author{Hubert Hua Kian Teo}
\affiliation{%
  \institution{Stanford University}
}
\email{hteo@stanford.edu}

\author{Conor McAvity}
\affiliation{%
  \institution{Stanford University}
}
\email{cmcavity@stanford.edu}

\author{Amit Levy}
\affiliation{%
  \institution{Princeton University}
}
\email{aalevy@cs.princeton.edu}

\author{Philip Levis}
\affiliation{%
  \institution{Stanford University}
}
\email{pal@cs.stanford.edu}

\renewcommand{\shortauthors}{H. Ayers et al.}

\begin{abstract}
  Over the past 10 years, low-power wireless networks have
  transitioned to supporting IPv6 connectivity through 6LoWPAN, a set
  of standards which specify how to aggressively compress IPv6 packets
  over low-power wireless links such as 802.15.4. 
  
  We find that different low-power IPv6 stacks are unable to communicate 
  using 6LoWPAN, and therefore IP, due to design tradeoffs between code size and energy efficiency. We argue
  that applying traditional
  protocol design principles to low-power networks is responsible for
  these failures, in part because receivers must accommodate
  a wide range of senders.

  Based on these findings, we propose three design principles for
  Internet protocols on low-power networks. These principles are based
  around the importance of providing flexible tradeoffs between code size
  and energy efficiency. We apply these principles to 6LoWPAN and show
  that the resulting design of the protocol provides developers a wide
  range of tradeoff points while allowing implementations with
  different choices to seamlessly communicate.
\end{abstract}


\maketitle

\section{Introduction}
Interoperability has been fundamental to the Internet's success. The
Internet Protocol (IP) allows devices with different software and
different link layers to communicate, while higher layer protocols
allow them to share a rich suite of networked applications. In the
decades of the Internet's evolution, we have accumulated and
benefitted from from a great deal of wisdom and guidance (e.g.,
Postel's Law) in how to design, specify, and implement robust,
interoperable protocols.

Over the past decade, the Internet has extended to tens of billions
of low-power, embedded systems such as sensor networks and the
Internet of Things. In particular, hundreds of
proprietary protocols have been replaced by 6LoWPAN, a standardized
format for IP networking on low-power wireless link layers such as
802.15.4~\cite{RFC6282} and Bluetooth Low Energy~\cite{RFC7668}.
Many embedded operating systems have adopted 6LoWPAN~\cite{tinyos,riot,mbedos,contiki-ng,contiki, lite-os, zephyr}
and every major protocol suite uses it ~\cite{zigbee,openthread}.

The increasing dominance of 6LowPAN would suggest a future in which
tens of billions of connected devices can communicate with each other.
In fact, devices today can
communicate with the broader Internet. However, in many cases they
cannot communicate \emph{with each other}. We find that no
pairing of the major implementations fully interoperates (Section~\ref{s:interop}).
6LoWPAN was created with the express purpose of
bringing interoperable IP networking to low power devices, yet two key features of the
protocol --- range extension via mesh networking of devices, and the convenience of
different vendors being able to share a gateway --- are largely impossible
over 10 years later.

Each of the openly available 6LoWPAN stacks -- most of which are
used in production -- implements a subset of the protocol.
Furthermore, each implementation includes compile-time flags to cut
out additional compression/decompression options. As a result, two devices might
both use 6LoWPAN, yet
be unable to exchange IP packets. This is especially problematic for
gateways, which \textit{need} to be able to talk to multiple implementations
to enable significant scaling of these devices in real world applications.
Because there is no mechanism to discover which protocol features a
particular device supports, gateways must either conservatively minimize
compression to ensure devices can receive its packets, or somehow determine supported features out-of-band
(e.g.\ only supporting devices from the same vendor).

This paper argues that the failure of 6LoWPAN interoperability stems from applying traditional
protocol design principles to low-power networks.  Low-power
protocols minimize energy consumption by compressing packets.
Squeezing every bit out of packet headers, however, requires many
different options and operating modes.  Principles such as Postel's
Law~\footnote{"Be liberal in what you accept, and conservative in what
  you send"~\cite{RFC1122}} imply that an implementation must be able
to receive all of a protocol's features, even if it never sends them.
However, code space is a tight constraint on most embedded systems. 
When application-critical code + networking code does not all fit on-device, implementations
opt to partially implement network protocols, and break interoperability
in the process.

Low-power Internet protocols need different design principles.  In
particular, low-power devices face a tension between code size and
energy efficiency.
Because application requirements vary,
there is no ``one size fits all'' choice for this tradeoff.  Instead,
individual applications must be able to decide how to balance
communication efficiency and code size.
Currently, when devices pick different points in this design space, they may fail to communicate.

To address this problem, we propose three design principles for low power
protocols:

\vspace{1ex}\noindent{\bf Capability spectrum:} a low-power protocol
specifies a linear spectrum of capabilities. Simpler implementations have fewer capabilities and save less energy, while fuller implementations have strictly more capabilities and are able save more energy. When two devices differ in capability
levels, communication can always fall back to the lower one.

\vspace{1ex}\noindent{\bf Capability discovery:} a low-power protocol
provides mechanisms to discover the capability of communicating devices. This discovery can be proactive (e.g., a device advertises
its capability level) or reactive (e.g., a device sends an error when
it receives a packet it cannot process).

\vspace{1ex}\noindent{\bf Explicit, finite bounds:} a low-power protocol specifies explicit,
finite bounds on how big a decompressed packet can be. Without explicit bounds, receiving every valid packet requires
allocating buffers too conservatively. In practice, implementations
allocate smaller buffers and silently drop packets they
cannot decompress.

\vspace{1ex}

We examine how these principles could be applied to a low power protocol and evaluate the code size overhead of capability discover. We find that applying these principles to 6LoWPAN promises interoperability across a wide range of device capabilities, while imposing a code size cost of less than 10\%. In particular, capability discovery requires an order of magnitude less code than the code size difference at the extremes of our capability spectrum.

%COAP. Given code size is
%such a tight constraint, one obvious concern is that adding capability
%discovery might itself add more code and so be the first feature that
%is excluded. We find that the cost of these principles is less than
%10\%. The code size difference between the simplest (only decompresses
%source addresses) and most full-featured 6LoWPAN implementation is
%just under 3kB, but the cost of capability discovery is, on average,
%280 bytes. This 10\% cost promises that low-power IP devices can
%communicate with IP.

\section{Background}
%Original intro to this section, pre-adding "Iot Connectivity" subsection. Could go back to it.

% In this work, we focus on protocols for applications intended to be very low power
% and low cost, such as long-life sensor nets or very small form factor IoT devices.
% Cellular IoT solutions like LTE-M or Nb-IoT are
% infeasible for these applications due to power and cost, but connectivity is still desirable.
% Devices in these low-power networks typically use ultra-low power, low-cost microcontrollers.
% The marginal cost of additional resources, in terms of both price and energy, is significant. At the same time, radio communication usually dominates the energy budget
% for networked applications: each transmitted bit consumes as much energy as hundreds or
% thousands of instructions.

In this work, we focus on protocols for applications intended to be very low power
and low cost, such as long-life sensor nets or very small form factor IoT devices.
Devices in these low-power networks typically use ultra-low power, low-cost microcontrollers.
The marginal cost of additional resources, in terms of both price and energy, is significant. At the same time, radio communication usually dominates the energy budget
for networked applications: each transmitted bit consumes as much energy as hundreds or
thousands of instructions.

\subsection{Low Power Protocols}

These embedded and low-power networked systems have historically been dominated by a
communication model based on ad-hoc custom designs. These systems
were vertical application silos: each manufacturer, or even each
application, defined its own protocol stack. Even in places where rich
application-level standards exist, such as Bluetooth, most devices
implement their own APIs and abstractions, precluding interoperability
or larger composition. 

In recent years, a number of standards have emerged to attempt to break out of these
vertical silos. 
Cellular IoT solutions like LTE-M or Nb-IoT offer broad coverage and high data rates, at the cost
of monthly service charges and support limited to more expensive and powerful embedded hardware ~\cite{nb-iot}. As a result, these solutions are outside the scope of this work.
Low-power wide-area networks (LPWANs) such as LoRaWAN and SigFox offer long range connectivity for
lower power/cost then cellular IoT, but force much lower data rates, and do not support IP
addressable nodes ~\cite{6low-lora-comp}.

But the Internet always wins. The communication and interoperability
IP provides is a goal in and of itself. It
allows systems to easily incorporate new services and applications,
and allows applications to build on all of the existing Internet systems.
Over the past decade, most major embedded operating systems and network
stacks have transitioned to using
6LoWPAN~\cite{tinyos,contiki,mbedos,contiki-ng,riot,openthread}, a standard protocol
to communicate IPv6 over low-power wireless link layers
such as IEEE 802.15.4 and Bluetooth~\cite{RFC4944, RFC6282, RFC7668}. 6LoWPAN primarily
specifies two things: compression of IPv6 headers and how to
fragment/re-assemble packets on link layers whose MTU is smaller than
the minimum 1280 bytes required by IPv6.

The core 6LoWPAN specifications are quite succinct: RFC6282 is 22
pages long. However, these 22 pages contain 19 different compression
mechanisms, covering only basic IPv6 data frames. There are
separate documents for neighbor discovery (ND) and other mechanisms.
The savings from these compressions schemes are significant: in the
best case, it can compress a 40-byte IPv6 header to two bytes: one
which specifies how to derive addresses and other fields, and one
containing the hop limit. For small payloads typical in low-power devices, 6LoWPAN can reduce the 
overhead of IPv6 from 400\% to 20\%.

\subsection{Application Specificity}

Even within this low power/low cost space, devices vary
significantly in capability and focus --- some devices are limited
primarily by power, others by cost, others by size, and others by
processor resources. Every combination of device and application has
a unique set of requirements for each of these parameters. For example,
some low-power applications may require the
battery last at least a year, while others may require the device used be small enough to attach to a bird
or go unnoticed in a piece of clothing. Other applications may need
on-device data processing operations that requires large libraries,
and still others may require RAM-hungry cryptographic primitives.

Despite this diversity of devices, the difficulties associated with
writing custom embedded firmware for an application means that a small
set of embedded operating systems are used for most embedded
applications. These embedded operating systems are expected to serve a
wide set of devices - and are unlikely to become popular if they
cannot accomplish this. Ultimately, this leads to a reality in which
limiting embedded operating system code size is of paramount
importance.

\subsection{Hardware Constraints}

\begin{table}[t]
\caption{Flash size varies widely across IoT platforms}\label{tab:platforms}
\begin{tabular}{lrr}
\toprule

\bf{IoT Platform}     & \bf{Code (kB)}   & \bf{Year} \\
\hline
Tmote Sky              & 48    & 2004 \\ %           & \$80\\
Zolertia Z1            & 92    & 2013 \\ %          & \$108\\
Atmel RZRaven          & 128   & 2007 \\ %          & \$120\\
TI CC2650              & 128   & 2015 \\ %          & \$41\\
NXP MKW40Z             & 160   & 2015 \\ %
SAMR21 XPro            & 256   & 2014 \\ %          & \$70\\
Nordic NRF52840 DK     & 512   & 2018 \\ %          & \$49\\
Arduino Due            & 512   & 2012 \\ %          & \$39\\
\bottomrule 
\end{tabular}
\end{table}

Table~\ref{tab:platforms} shows a variety of older and more recent
low-power platforms. Modern microcontrollers typically have 128-512 kB
of code flash.  This limited space holds the operating system,
networking software, sensor drivers, storage abstractions, and
application code. Chips have limited flash because the type of flash
used (NOR) is space-intensive: a chip with less flash is smaller and
so more can be made on a single silicon die. NOR flash is used
because, unlike much denser NAND flash, NOR allows random access,
which has better energy properties for instruction streams.

Microcontrollers have limited flash and applications struggle with
these limits. Given the expectation that these systems will run the
same application for years, applications do not leave code space
unused. To help accomodate these tight tolerances, embedded operating systems have a wide range of compile-time
flags to include or exclude parts of the system or networking
stack.~\cite{contiki,mbedos,contiki-ng} Some systems take an even more
extreme approach, dynamically generating the minimum code to compile
from the application itself.~\cite{tinyos} Moreover, chip vendors often allow application
developers to select their particular system/application size tradeoff
by providing multiple firmware images for the same chip, with different
levels of functionality. Nordic Semiconductor, for example, provides
three different SoftDevices for their Bluetooth Low Energy microcontrollers,
from the minimal S112 that only supports the peripheral role, to the fully featured
S132 which supports both peripheral and central roles with up to twenty
concurrent connections.

\section{Low-Power IP Today} \label{s:interop}
\begin{quote}
  {\it ``The Working Group will generate the necessary documents to ensure
interoperable implementations of 6LoWPAN networks''}

  \begin{flushright}
    \tiny{--- 6LoWPAN Working Group Charter~\cite{ietf-6lowpan-charter}}
  \end{flushright}
\end{quote}

This section gives an overview of 6LoWPAN and its implementations.  It
finds that no implementation today implements the entire protocol and that
these implementation holes are non-uniform: every pair of
implementations succeeds to receive packets in some cases but
deterministically fails to receive packets in others. Examination of
the source code finds that these problems are due to concerns with
code size. Experiences with a new, clean implementation verify these
concerns.

\begin{figure}
\centering
\includegraphics[width=3in]{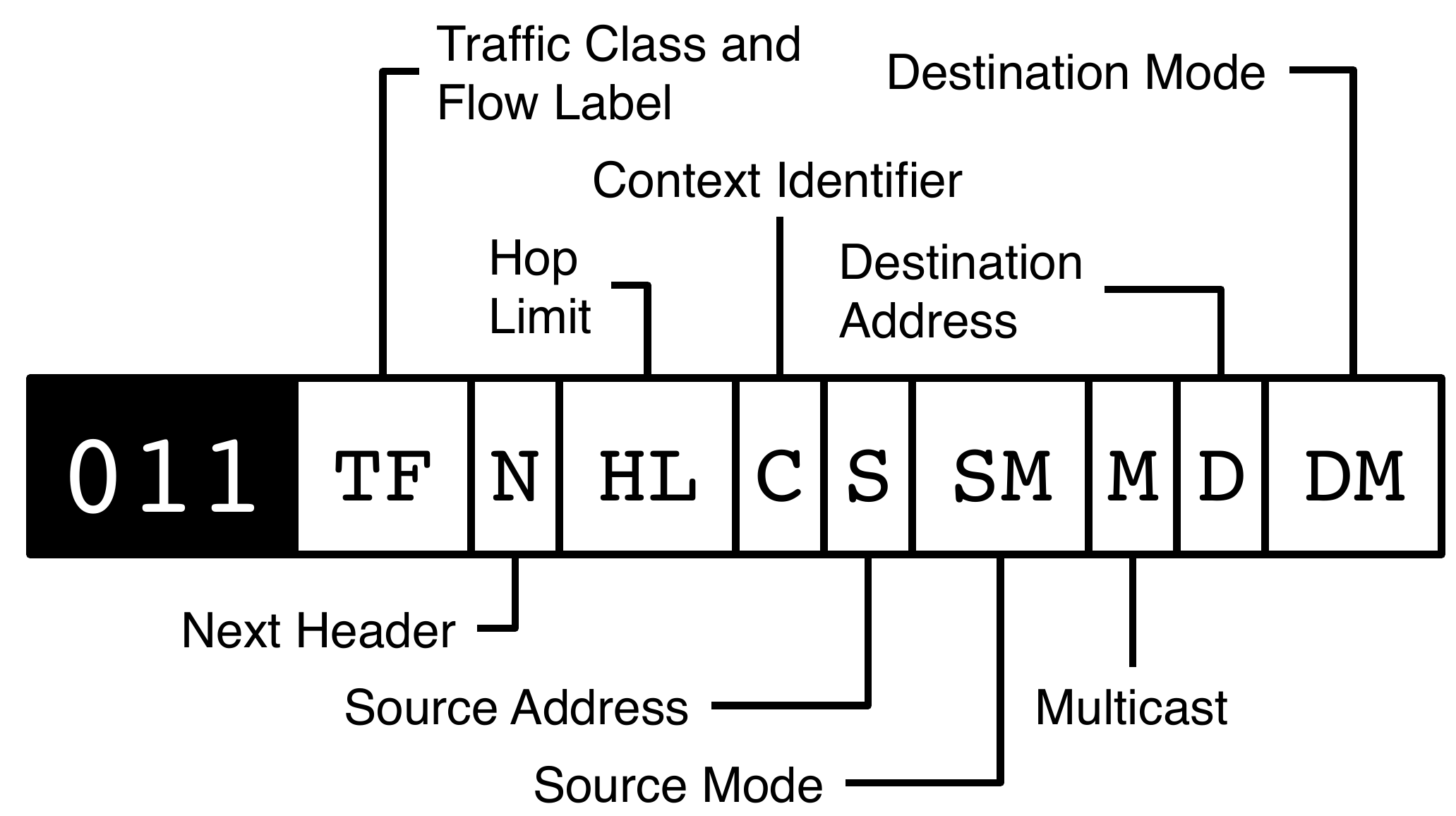}
\caption{The 6LoWPAN IPHC header compression encoding. In some cases,
  6LoWPAN can compress a 40-byte IPv6 header to these two bytes. The
  address bits denote whether stateful or stateless compression is
  used, while the address mode bits denote what portion of the address
  (128, 64, 16, or 0 bits) is included it the header or derived from
  other information.}
\label{fig:6lo-header}
\end{figure}

\subsection{6LoWPAN Summary}

\begin{table*}[t]
\caption{6LoWPAN Interoperability Matrix. No implementation can receive all viable messages in the 6LoWPAN specification, and no stack chooses to elide the same features, with each instead opting for a combination
that works best for its particular resource requirements.}\label{tab:a}
\begin{tabular}{lcccccc}
\toprule
\bf{Feature}                                                    & \multicolumn{5}{c}{ \bf{Stack} }                                     \\
\hline
                                          & \bf{Contiki} & \bf{Contiki-NG} & \bf{OpenThread}  & \bf{Riot}    & \bf{Arm Mbed} & \bf{TinyOS}        \\
\hline
Uncompressed IPv6                         & \checkmark & \checkmark   &             & \checkmark     & \checkmark      & \checkmark           \\
6LoWPAN Fragmentation                     & \checkmark & \checkmark   & \checkmark         & \checkmark     & \checkmark      & \checkmark             \\
1280 byte packets                         & \checkmark & \checkmark   & \checkmark         & \checkmark     & \checkmark      & \checkmark            \\
Dispatch\_IPHC header prefix              & \checkmark & \checkmark   & \checkmark         & \checkmark     & \checkmark      & \checkmark             \\
IPv6 Stateless Address Compression             & \checkmark & \checkmark   & \checkmark         & \checkmark     & \checkmark      & \checkmark            \\
Stateless multicast address compression   & \checkmark & \checkmark   & \checkmark         & \checkmark     & \checkmark  	& \checkmark            \\
802.15.4 16 bit short address support     &            & \checkmark   & \checkmark         & \checkmark     & \checkmark      & \checkmark            \\
IPv6 Address Autoconfiguration            & \checkmark & \checkmark   & \checkmark         & \checkmark     & \checkmark      & \checkmark            \\
IPv6 Stateful Address Compression              & \checkmark & \checkmark   & \checkmark         & \checkmark     & \checkmark      & \checkmark             \\
IPv6 Stateful multicast address compression    &            &              & \checkmark         & \checkmark     & \checkmark      &                \\
IPv6 Traffic Class and Flow label compression  & \checkmark & \checkmark   & \checkmark         & \checkmark     & \checkmark      & \checkmark           \\
IPv6 NH Compression: Ipv6 (tunneled)      &            &              & \checkmark         &         & \checkmark      & \checkmark             \\
IPv6 NH Compression: UDP                  & \checkmark & \checkmark   & \checkmark         & \checkmark     & \checkmark      & \checkmark            \\
UDP port compression                      & \checkmark & \checkmark   & \checkmark         & \checkmark     & \checkmark      & \checkmark            \\
UDP checksum elision                      &            &              &             &         &          & \checkmark            \\
Compression + headers past first fragment    &            &              &             & \checkmark     & \checkmark      &                \\
Compression of IPv6 Extension Headers     &            & \textasciitilde & \textasciitilde \tablefootnote{Contiki-NG and OpenThread do not support compression of the mobility header}     &         & \checkmark      & \checkmark    \\
Mesh Header                               &            &              & \checkmark         &         & \checkmark      & \textasciitilde \tablefootnote{TinyOS can receive packets which use the mesh header, but will never forward such packets or send packets using this header, effectively preventing TinyOS devices from participating in route-under networks}         \\
Broadcast Header                          &            &             &             &         &          & \checkmark           \\
Regular IPv6 ND                           & \checkmark & \checkmark         &             & \checkmark     & \checkmark      & \textasciitilde \tablefootnote{TinyOS supports only some of RFC 4861}        \\
RFC 6775 6LoWPAN ND                       &            &             &             & \checkmark     & \checkmark      &                \\
RFC 7400 Generic Header Compression       &            &             &             &         &          &                \\
\bottomrule
\textasciitilde\ = Partial Support
\end{tabular}
\end{table*}

6LoWPAN is comprised of 3 core documents - RFC 4944~\cite{RFC4944},
RFC 6282~\cite{RFC6282}, and RFC 6775~\cite{RFC6775}. 6LoWPAN defines
new header types to compress and structure IPv6 packets over low-power
link layers.  Because link layers have different communication models
and address formats, each link layer has its own
specification.~\cite{RFC6282,RFC7668,RFC7428,RFC8105,RFC8163} In this
paper we focus on 802.15.4, since it was the original driver for
6LoWPAN and its dominant use case. 6LoWPAN solves three major
problems: header compression, stateless address autoconfiguration, and
sub-IP fragmentation.

A standard IPv6 header is 40 bytes. Low-power devices, however, have
small MTUs (Bluetooth Low Energy, for example is 27 bytes) and often
send small data payloads (e.g., 10 bytes). 6LoWPAN therefore provides
mechanisms to heavily compress IPv6 packets.  Because addresses
dominate the header, there are context-based and context-free
compression schemes for unicast and multicast IPv6 addresses, as well
as cases in which other fields within IP and UDP headers may be
compressed.

Figure~\ref{fig:6lo-header} shows the layout of the two-byte header
compression field. For example, the Traffic Class and Flow Label
fields can be elided, while the Hop Limit field can be compressed if
it is one of several common values and certain UDP ports can be
compressed.  In the most extreme case, 6LoWPAN can compress the
40~byte header to 2 bytes; for a 10 byte packet, over Bluetooth Low
Energy, this is the difference between a 400\% overhead and sub-IP
fragmentation or a 20\% overhead and no fragmentation.

This compression is tightly entwined with the second problem,
stateless address autoconfiguration. Each link layer defines how to
map unicast and multicast IPv6 addresses from and to its link layer
addresses.

The third major problem 6LoWPAN solves is fragmentation. IPv6 requires
that a link layer support 1280 byte packets without IP fragmentation.
Because low-power link layers often have small MTUs (e.g., 802.15.4
has an MTU of 127 bytes), 6LoWPAN defines how to fragment and
reassemble larger-than-MTU packets using a fragmentation header.

In addition to these three core problems, the protocol covers edge
cases related to fragmentation and compression, such as restricting
header compression to the first fragment of a fragmented packet.
Finally, RFC 6775~\cite{RFC6775} describes 6LoWPAN optimizations for
IPv6 Neighbor Discovery.

\subsection{Feature Fail}\label{ss:Incomplete}

6LoWPAN gives senders a broad range of compression options to use. For
example, a sender can choose to simply not compress a packet at all,
e.g.  a full 40-byte IPv6 header can follow the 2-byte 6LoWPAN header.
A receiver, however, has much less flexibility: it must be able to
receive and process any valid compression it receives. Every 6LoWPAN
receiver is required to parse multiple 6LoWPAN header types, including
the mesh addressing header, broadcast header, fragmentation header,
and compression headers. 

\begin{table}[t]
\caption{Hardware/Software used for code size measurements and
IPv6 communication tests}\label{tab:c}
\begin{tabular}{lp{2.6cm}p{2.9cm}}
\toprule
Stack      & Commit Hash                              & Device                 \\
\hline
Contiki    & bc2e445817aa546c & CC2650 LaunchXL        \\

Contiki-NG & 7b076e4af14b2259 & CC2650 LaunchXL        \\

OpenThread & 4e92a737201b2001 & Nordic NRF52840    \\

Riot       & 3cce9e7bd292d264 & SAM R21 X-Pro   \\

Arm Mbed   & 4e92a737201b2001 & N/A                   \\

TinyOS     & 4d347c10e9006a92 & Atmel SAM4L            \\
\bottomrule
\end{tabular}
\end{table}

Table~\ref{tab:a} shows the features supported by 6 major open-source
6LoWPAN stacks. Some, such as TinyOS, are mostly developed and used in
academia. Others, such as ARM Mbed and Nest's OpenThread, are
developed and supported commercially. Contiki and Contiki-NG sit
somewhere in the middle, having both significant academic and
commercial use. Riot is an open-source operating system with hundreds
of contributors for industrial, academic, and hobbyist use.

There are significant mismatches in feature support between stacks.
These mismatches lead to deterministic cases when IP communication
fails. We verified these failures by modifying existing network
applications and testing them on hardware, using Wireshark to verify
packets were compressed and formatted as we expected and that
receivers would fail to receive packets. Every implementation pair fails in some
fashion:

\begin{itemize}
\item \textbf{Contiki $\rightarrow$ OpenThread} : Contiki generated message using uncompressed IPv6
\item \textbf{Contiki-NG $\rightarrow$ Contiki} : Contiki-NG generated message with compressed IPv6 extension headers
\item \textbf{Riot $\rightarrow$ Contiki}: Riot generated message using stateful multicast address compression
\item \textbf{Mbed $\rightarrow$ Contiki}: Mbed generated message using compressed, tunneled IPv6

%-- This is made possible by line 472 in iphc\_compress.c which will recursively
%compress any next header, including the Ipv6 header.
\item \textbf{TinyOS $\rightarrow$ Contiki}: TinyOS generates messages containing a compressed IPv6 mobility header

%-- Generating such a packet is made possible by line 143 in lb6lowpan.c
\item \textbf{OpenThread $\rightarrow$ Riot}: OpenThread generates messages containing any
of the IPv6 extension headers, which the OpenThread stack automatically compresses
\item \textbf{Mbed $\rightarrow$ OpenThread}: Mbed generates IPv6 packet containing the IPv6 mobility header
\item \textbf{OpenThread $\rightarrow$ TinyOS}: OpenThread generates message for which the destination address
is compressed using stateful multicast compression
\item \textbf{Mbed $\rightarrow$ Riot}: Mbed generates IPv6 message containing any compressed next header other
than the UDP header
\item \textbf{Riot $\rightarrow$ TinyOS}: Riot generates message for which the destination address is
compressed using stateful multicast compression
\item \textbf{Mbed $\rightarrow$ TinyOS}: Mbed generates Neighbor Discovery message using the 6LoWPAN
context option as specified in RFC 6775.

\end{itemize}

This is a non-exhaustive listing -- in most pairings there are many
packet formats which would cause IP to fail.  Table~\ref{tab:c} lists
the exact software/hardware combinations we used for our code analysis
and harware tests.

In addition to the IPv6 header itself, 6LoWPAN receivers are expected
to decompress some next headers as well as destination options. The
ability to compress next headers is critical for low-power routing
protocols such as RPL~\cite{RFC6550}, which often tunnel IPv6 inside
IPv6 to manage ICMP errors.~\footnote{If a gateway adds source routing
  headers and these cause ICMP errors later, the errors should go
  to the gateway, not the source host. So gateways tunnel IPv6 packets
  inside IPv6 packets they source.} Every stack, however, fails to
decompress tunneled IPv6 packets, while Contiki and OpenThread fail to
decompress any extension headers.

\subsection{Constant Disagreement}

In addition to the feature mismatches in Table~\ref{tab:a}, IP
communication between different stacks can fail due to disagreement on
certain constants. These disagreements generally arise out of concerns
for RAM use by one of the participants. One example where this
can lead to problems is when stacks make assumptions about
the maximum amount of decompression which is possible from
a single link layer frame. 6LoWPAN's extensive compression means that a
decompressed packet can be much larger than what is received over the
air. Further, because of 6LoWPAN fragmentation, fragments of a packet
can arrive out of order. To maintain layering, and prevent 
interleaved or stray packets from 
starving the single MSS size receive buffer present
on most devices, most stacks decompress initial 6LoWPAN fragments 
into a slightly extended link layer size buffer.
One implementation decision a stack must make is how much RAM to
allocate for each of these extended frame buffers. Given that
most devices in this space do not have the option of dynamic
allocation, this can be an impactful decision.

As a result, every stack we analyzed imposes a limit on the amount of header
decompression possible in a received packet.  The maximum amount of
header decompression allowed by the 6LoWPAN specification is about 1200
bytes, basically, if an entire Maximum Segment Size (MSS) IPv6 packet
was sent containing only compressed headers. Stacks place much lower
limits to avoid a requirement for multiple IPv6 size buffers which
would mostly sit empty. For example, Contiki's 38
byte limit is exceeded by any packet with a maximally compressed IP
header and any UDP compression. Contiki merely seems to have chosen
a 38 byte limit because the limited Contiki stack will not compress frames
by more than that amount. We verified that some stacks send
packets with more header compression than this limit, 
causing IP communication to fail.

\subsection{Why?}

IP communication routinely fails in low-power networks, despite the
presence of succinct standards (RFC6282 is 22 pages) explicitly
designed for low-power embedded devices.  Furthermore, this failure is
silent. For example, if a Contiki device tries to forward an
uncompressed IPv6 packet to an OpenThread device, the OpenThread
device will drop the packet. 

Examining the stacks and their implementations, we find a single, common cause:
concerns about code (and to a lesser degree, RAM) size.

The tensions between RAM and code size has always been a delicate
dance on low power embedded platforms. For example,
designing for the particular RAM/code size split of one architecture
can be problematic for others: TinyOS service APIs had to change
dramatically over time to accomodate devices with different ROM/RAM
splits than the mica platform for which it was originally
developed~\cite{TinyOSDev}.

In low power networking, another important ``slider'' exists---the
tradeoff between code size and network protocol efficiency.
Techniques such as advanced MAC and physical layers, and tracking the
state of a network can reduce packet sizes and, thus, radio energy
consumption.  However, these techniques require more complex
implementations. Increased code size, however, requires more expensive
and power hungry micrcontrollers.

6LoWPAN takes great efforts to save energy through IPv6 header
compression. The result, however, is that implementations take up a
large fraction of the code space on small devices.  Analyzing the
comments and documentation of each stack, we found that code size
concerns motivated features elision. Mbed, Riot, and Contiki take this
one step further, including compile-time flags to remove features.

\begin{figure*}
\includegraphics[width=0.9\textwidth]{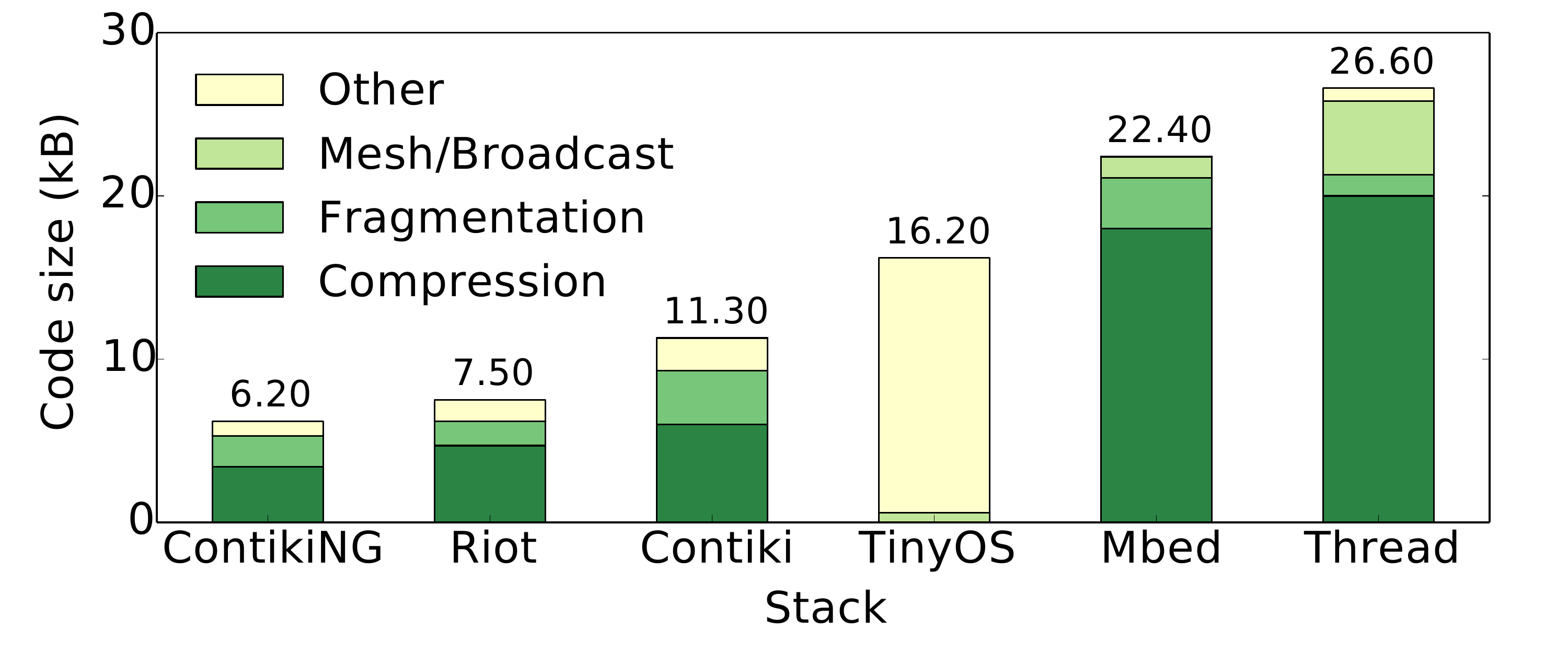}
\caption{6LoWPAN Stack Code Size for 6 open source stacks. The code size varies
by over a factor of 4, in part due to different feature sets. In all cases,
compression dominated the code requirements of each stack. In some cases,
the size of the 6LoWPAN implementation is larger than the rest of the IPv6 stack. TinyOS's
whole-program optimization model precluded separating out subcomponents
of its implementation.}
\label{fig:code-size}
\end{figure*}

Figure~\ref{fig:code-size} shows the
code size of each of the six implementations broken into independent
compression, fragmentation, mesh and broadcast headers, the rest of
6LoWPAN and the rest of the IPv6 networking stack (physical layer
drivers, IPv6, UDP, and ICMP).  
Compression dominates the code
size of 6LoWPAN implementations, and in several cases 6LoWPAN's size
is comparable to the whole rest of the IPv6 stack. Contiki, and
Contiki NG implementations are significantly smaller than the others in
part because they elide significant and complex features.  The ARM
Mbed IPv6 stack, for example, uses 45kB of flash. This is nearly 1/3
of the available space on a CC2650, just for IPv6: it does not include
storage, sensors, the operating system kernel, cryptography, higher
layer protocols, signal processing, or applications.

%the Contiki, Contiki-NG, and Riot build systems to remove particular 6LoWPAN functionality at
%compile-time. For OpenThread, TinyOS, and Mbed, which do not have such options, we
%manually analyzed binary build products using the \texttt{readelf} utility from
%the GCC toolchain to isolate functions associated with each aspect of 6LoWPAN.

%These results likely overestimate the overhead of fragmentation and
%underestimate the overhead of certain kinds of compression since some of the
%complexity of compression is born on the fragmentation logic. Moreover, for
%OpenThread and Arm Mbed, where we had to examine binaries manually, we expect
%the results underestimate the overhead of all 6LoWPAN components since we
%only counted procedures which unambiguously implemented particular
%functionality, though some of the complexity is implemented in other portions of
%the stack. In summary:

%% What should the reader take away from these results, as a bullet list for
%% now
%\begin{itemize}
%    \item 6LoWPAN stack devolopers were concerned with 
%        processor resource requirements of the protocol.
%	\item Fragmentation, the only portion of 6LoWPAN that's strictly
%		\emph{necessary} for sending IPv6 packets, consumes
%		significantly less ROM than compression.
%	\item Implementations with more complete adherence to compression
%		specification generally consume more code for compression
%\end{itemize}

\subsection{Investigating Ourselves}

Can a careful developer avoid these pitfalls and implement a lean,
fully-featured stack? To answer this question, we implemented our own
full 6LoWPAN stack. Our 6LoWPAN stack is written in Rust, for Tock, a
secure embedded operating system that provides a safe multiprogramming
environment for microcontrollers.~\cite{tock}

Our experiences support the comments and documentation of the other
stacks.  Furthermore, we discovered parts of it that require
particularly surprisingly complex and large code to properly handle.
For example, tunneling IPv6 inside a compressed packet requires the
interior headers be compressed as well.  This requires the
decompression library to support recursive invocation, which increases
minimum stack sizes and makes tracking buffer offsets during
decompression more difficult.  Refusing to support tunneled IPv6
packets (e.g., Contiki) greatly simplifies the code. Another example
is the fact that headers in the first 6LoWPAN sub-IP fragment must be
compressed, while headers in subsequent fragments must not be
compressed. Given that low-power link layer have variable length
headers, correctly determining exactly where to fragment and what
should be compressed requires complex interactions between layers of
the stack. The exercise of implementing a 6LoWPAN stack from the
ground up afirmed the observation that there is a tension between
code size and energy efficiency in low power protocol design.

\subsection{Why Does This Matter?}

For each unique
6LoWPAN implementation, there is presumably an accompanying border router implementation
that always successfully interoperates with this device, and pairing the two will make that device available
to the broader Internet. But this status-quo model of connectivity forces vertical integration
and fails to meet the original design goals of 6LoWPAN, for two reasons.

First, the current situation is such that a 6LoWPAN gateway cannot know how much compression is allowable
for IP packets intended for different nodes, unless that gateway communicates only with devices produced by the same vendor.
As a result, for a coverage area containing devices produced by 5 disparate vendors, 5 6LoWPAN
gateways are required. Each device can communicate only with the gateway produced by its
vendor, or directly with other devices by the same vendor. If not for the feature mismatch we have demonstrated, a single gateway would suffice,
an approach which has substantial benefits from the perspectives of usability, cost, and wireless efficiency. Secondly, the current situation
significantly limits the potential range extension made possible via mesh topologies. Most existing 6LoWPAN meshes rely on
route-over mesh routing at the network layer, which requires that each node can at least partially decompress and recompress IP headers when forwarding frames sent by other nodes.
Even mesh-under routing is not a reliable option, as implementation of the mesh header is not ubiquitous as shown in table~\ref{tab:a}.
Ubiquitous direct connectivity between 6LoWPAN nodes is crucial in realizing the range extension and cost/power savings promised by 6LoWPAN mesh topologies.

\section{Traditional Principles in the Low-Power Domain} %TODO: Better title
Connectivity through interoperability is a key premise of the Internet:
Following this premise has enabled it tremendous success in connecting
a diverse and varied collection of 
different networks and software stacks into a single, cohesive whole. 
Over the past 45 years, we have coalesced
on a small number of principles that lead to protocols which
are simultaneously well-defined enough to support many implementations yet
flexible enough to evolve and improve as networks change. Principles such
as layering and encapsulation support composing protocols in new ways
(e.g., tunneling), while the end-to-end principle~\cite{end-to-end} allows us to build
a robust network out of an enormous and complex collection of unreliable parts. 
The simplicity principle~\cite{RFC3439} discourages complexity in protocols,
and is used to support claims like
``optimization can also be considered harmful. In particular, optimization introduces
complexity, and as well as introducing tighter coupling between components and layers''~\cite{doyle}.
There are great
reasons for traditional Internet principles, and they have generally 
served us well.

The robustness principle is one such principle. The robustness
principle asserts that implementations should make no
assumptions on packets they receive: bugs, transmission errors, and
even memory corruption can cause a device to receive arbitrarily
formatted packets.  An implementation
must also be ready to receive and
properly discard packets that are arbitrarily malformed. As one RFC
puts it, ``In general, it is best to assume that the network is filled
with malevolent entities that will send in packets designed to have
the worst possible effect.''~\cite{RFC1122}

The robustness principle also asserts that an implementation must be
ready to receive any and all properly formatted packets. This aspect
of the principle is often attributed to John Postel as Postel's Law,
first written down in
the initial specification of IPv4: ``In general, an implementation
should be conservative in its sending behavior, and liberal in its
receiving behavior.'' While some have recently argued that instead
protocols should be maximally strict~\cite{postel-was-wrong}, even this
restriction of Postel's Law assumes that an implementation properly
handles the full protocol.

Protocols often have optional features (``MAY, SHOULD, and OPTIONAL'' in
RFC language). These features are often accompanied by explicit instructions
for how receivers should act in the presence of these optional features, or
lack thereof. For example, RFC6282 says that
interface identifier bits may be elided if they match the link layer address
(effectively deriving part of the IPv6 address from the link layer address);
this means that a sender can but does not have to elide in this way.
Implicitly, due to Postel's Law, however, a receiver needs to be able to handle
either case. This scenario creates an interesting asymmetry, where
it can be easy to write sender code with reduced complexity but difficult 
to write receiver code that does the same. This asymmetry leads to a situation in
which meaningful reductions in complexity/code size are difficult to realize while
maintaining interoperability and following the robustness principle.

A similar example is
UDP checksum elision, one of the least supported features in Table~\ref{tab:a}. At the
6lo working group meeting at IETF 102, one of the authors of RFC 6282
noted that UDP checksum elision was including at the request of a vendor who wanted
to use 6LoWPAN and had applications that required this feature~\cite{ietf102}.
Unfortunately, without capability discovery or error signaling, the existence of this feature
in the specification implicitly adds a requirement that
all receivers support decompression of packets formatted in this manner. This is an undesirable situation.

In low power networks, following the principle that receivers should
accommodate every sender is expensive. Senders
have the choice of what features to implement, but receivers have to
implement every one and handle every combination.
Following the simplicity principle is hard when writing low-power protocols, as optimizations are crucial
in the low-power space. But protocols need to make it possible for individual vendors to use optimizations without
breaking interoperability by imposing code size costs on all implementations.

We need to think about low-power protocols differently. They need
new principles to help guide their design. These principles need
to embrace the fact that there is no ``one size fits all'' design,
while defining how devices choosing different design points interoperate.
Flexibility needs to exist not only for senders, but also for receivers,
without harming interoperability.
The next section describes three principles that achieve this goal,
and following sections describe an application of them to the 6LoWPAN
protocol suite.

\section{Three Principles}
\begin{quote}
\it{``A good analogy for the development of the Internet is that of
   constantly renewing the individual streets and buildings of a city,
   rather than razing the city and rebuilding it.''}

  \begin{flushright}
    \tiny{--- RFC 1958, Architectural Principles of the Internet~\cite{RFC1958}}
  \end{flushright}
\end{quote}

In light of the conflicts between traditional Internet principles
and interoperability on low power devices, this section describes 
three protocol design principles which, when working on low-power 
protocols, should be closely observed. These principles are
of a different nature than many traditional Internet principles, but in this space
are absolutely necessary in ensuring
interoperable implementations. In the next section, we show how to apply each to 6LoWPAN.

\subsection{Principle 1: Capability Spectrum}

A low power protocol should be implementable on devices which are at
the low end of code and RAM resources. Rather than require every
device pay the potential energy costs of fewer optimizations, a
protocol should support a linear spectrum of device capabilities. 

At first glance, this principle may seem to suggest an approach
already applied by traditional Internet protocols --- the IP~\cite{RFC0791}and TCP~\cite{RFC0793}
specifications provide optional fields which can be used by endpoints
at their leisure; many non-standard HTTP headers will be ignored unless
both client and server support them; TLS ciphersuite support is often
asymmetrical. In reality, the capability spectrum
necessary for a low power Internet protocol is distinctly
different from any of these examples. For each of these examples, no linear
spectrum exists --- support, or lack thereof, for any particular capability in
each of those examples is generally unrelated to support for other options
in the set. Checking for support of any of these options requires
explicit enumeration of each. Such a design generally
precludes the possibility of effective compression of such options,
as the options themselves would become difficult to parse without additional
information. Further, a non-linear spectrum means that storing capability
information for neighbors requires storing
details about every single optional aspect of the protocol for each neighbor,
or requires re-discovering capabilities on every exchange. 

Low power protocols, on the other hand, require simple spectrums,
such that the spectrum truly enables simpler
implementations from the perspective of both RAM and code size.
A low power protocol should define a capability spectrum with a 
clear ordering via which especially resource
constrained devices can reduce code size or RAM use by eliding
features. Such a spectrum makes a protocol usable by extremely low
resource devices without forcing more resourceful devices to
communicate inefficiently.

This capability spectrum should be a linear scale. For a device to
support capability level $N$, it must also support all lower
capability levels. More complex configuration approaches (e.g., a set
of independent options) would allow for a particular application or
implementation to be more efficient, picking the features that give
the most benefit at the least complexity cost. However, this sort of
optimization then makes interoperability more difficult, as two
devices must negotiate which features to use.

\subsection{Principle 2: Capability Discovery}

The second principle follows immediately from the first: if two
implementations may have different capability levels, there should be
an explicit mechanism by which two devices can efficiently determine
what level to use when they communicate.

Once again, we would like to emphasize that the capability negotiation
we propose as appropriate for this space is different from capability
discovery mechanisms built for traditional systems, such as IP Path 
MTU discovery or the Link Layer Discovery Protocol (LLDP). 
Though the MTU of a given path is by definition a linear scale,
IP Path MTU discovery relies on continual probing until an acceptable
value is discovered. The energy overhead of such network probing is unacceptable
in most low power environments. Given this, \textit{capability discovery in
low power networks should require no more than one failure between
any two neighbors}, even if this slightly increases the overhead
of handling a single error. Further, assumptions for
traditional systems that prohibit storing per-endpoint state do not
hold for these systems. On low power networks, it is feasible
to store state for a small number of neighbors --- most low power nodes communicate
with only a few neighbors at a time, so storing state can significantly
reduce the amount of radio energy needed for communication. 
Further, the code size cost of storing state is small compared
to the cost of complex compression mechanisms.

Similarly, LLDP requires regular capability advertisements at a fixed interval which
contain detailed capability information. 
This regular, unique, and proactive capability discovery is impractical in low power networks,
where radio energy is at a premium. Instead, low power networks should incorporate
any proactive capability discovery mechanisms into other baseline communication required
for tasks such as neighbor discovery or route maintenance. Beyond that, these networks
should rely on reactive discovery and small amounts of stored state.

In a low power network with capability discovery, if two devices
wish to communicate, they default to the lower of
their supported capability levels. For example, suppose a TinyOS
device supports level 2 and a Contiki device supports level 4; Contiki
must operate at level 2 when communicating with the TinyOS device.
This requires keeping only a few bits of state for each neighbor a node
may want to communicate with. Also, note that this state is per-hop; for a layer 3
protocol like IP, it is stored for link-layer neighbors (not IP
endpoints) and so does not require knowledge of the whole network. This
is the case because route-over topologies commonly used in low power IP
networks frequently involve decompression and re-compression at each hop
to enable forwarding.
One offshoot of this principle is that it requires implementations have
symmetric capabilities for send and receive -- if the two
nodes will always default to the lower of their support capability levels,
no benefits can be realized from an asymmetric implementation.

\subsection{Principle 3: Explicit and Finite Bounds}

Protocols should specify explicit and reasonable bounds on recursive or variable features
so implementations can bound RAM use. These bounds have two benefits.
First, they allow implementations to safely limit their RAM use without
silent interoperability failures. E.g., today, if an mbed device sends
a 6lowpan packet whose compression is greater than 38 bytes to a
Contiki device, Contiki will silently drop the packet. Second, it
ensures that capability discovery is sufficient for interoperability.

The idea of imposing bounds is, on its own, not unique to this space. TCP enforces
a finite limit of 40 bytes for TCP options which may be appended to the TCP header, as
does IPv4. DHCP allows for the communication of maximum DHCP message sizes.
In the space of low power Internet protocols, however, this idea \textit{must be
pervasive.}
What is unique to space of low power embedded networking is the importance of 
compression at all layers of the stack, which leads to scenarios in which,
without reasonable, finite bounds, it can be difficult for implementations to
impose appropriate maximum buffer sizes. Many low power systems lack dynamic allocation,
but do allow for the storage of multiple datagrams of various types. The ability to
precisely set the buffer sizes required to hold these packets is an important concern,
and undefined bounds on decompression or variable options can lead to significant RAM waste
summed across protocols. Rather than forcing implementers to make difficult decisions
between interoperability and processor resources, low power protocols must recognize
that the energy gains of unbounded option sets are useful in a tiny set of cases. Further,
while recursive compression mechanisms can be tempting to allow for maximal packet
size reductions, implementing these recursive mechanisms can require tail
recursion on the stack, which is problematic for low resource devices, and
these recursive mechanisms are rarely used in practice.
RAM resources, on the other hand, are an omnipresent concern.

Reasonable, finite bounds must exist in every scenario for low power
Internet protocols. Notably. the original designers of a specification may not know exactly what
these values should be. This is not a new problem: TCP congestion
control, for example, had to specify initial congestion window values.
In this space, bounds should initially be very conservative. Over time, if
increasing resources or knowledge suggests they should grow, then
future devices will have the onus of using fewer resources to
interoperate with earlier ones. The capability spectrum defined in
the previous two principles can be helpful in this regard.

\section{A Principled 6LoWPAN} \label{s:appl}
This section proposes how to apply the three principles in the
previous section to 6LoWPAN through specific modifications to the protocol. 
These modifications ensure that two 6LoWPAN devices can
communicate even if they choose different code size/energy efficiency
tradeoffs. We refer to this modified protocol as Principled 6LoWPAN (P6LoWPAN).

This application of our principles is not intended as a suggestion that
these changes should be made immediately to 6LoWPAN, as modifying an established protocol
is a complex task very different from constructing new protocols. Rather, this application
is a means by which we can evaluate how these principles are useful toward developing
other Internet protocols in this space.

\subsection{Principle 1: Capability Spectrum}

We propose replacing the large collection of ``MUST'' requirements ---
the features in Table~\ref{tab:a}---into 6 levels of functionality.
These ``Capability Levels'' are depicted in Table~\ref{tab:cap_spec}

\begin{table*}[t]
\caption{Capability Spectrum}\label{tab:cap_spec}
\begin{tabular}{lp{65mm}p{85mm}}
\toprule
\textbf{Capability}      & \textbf{Basic Description}                              & \textbf{Added Features}                 \\
\hline
Level 0    & Uncompressed IPv6 + ability to send ICMP errors &     \begin{itemize}
        	\item Uncompressed IPv6
        	\item 6LoWPAN Fragmentation and the Fragment Header
            \item 1280 Byte Packets
            \item Stateless decompression of source addresses
        \end{itemize}   \\
\hline

Level 1    & IPv6 Compression Basics + Stateless Address Compression &       \begin{itemize}
        	\item Support for the Dispatch\_IPHC Header Prefix
        	\item Correctly handle elision of IPv6 length and version
        	\item Stateless compression of all unicast addresses
        	\item Stateless compression of multicast addresses
            \item Compression even when 16 bit addresses are used at the link layer
            \item IPv6 address autoconfiguration
        \end{itemize} \\
\hline
Level 2    & Stateful IPv6 Address Compression &   \begin{itemize}
        	\item Stateful compression of unicast addresses
        	\item Stateful compression of multicast addresses
        \end{itemize}  \\
\hline
Level 3       & IPv6 Traffic Class and Flow Label Compression &   \begin{itemize}
        	\item Traffic Class compression
            \item Flow Label Compression
            \item Hop Limit Compression
        \end{itemize}  \\
\hline
Level 4   & IPv6 and UDP Next Header Compression + UDP Port Compression & \begin{itemize}
        	\item Handle Tunneled IPv6 correctly
        	\item Handle the compression of the UDP Next Header
            \item Correctly handle elision of the UDP length field
            \item Correctly handle the compression of UDP ports
            \item Correctly handle messages for which headers go on 
            longer than the first fragment, and the headers in the first
            fragment are compressed.
        \end{itemize}                   \\
\hline
Level 5     & Entire Specification &    \begin{itemize}
        	\item Support the broadcast header and the mesh header as 
           	described in RFC 4944
        	\item Support compression of all IPv6 Extension headers
        \end{itemize}         \\
\bottomrule
\end{tabular}
\end{table*}

These levels prioritize features which provide the greatest energy
savings per byte of added code size, based off of our
measurements of code size, the number of bits saved by each additional compression
mechanism, and our observations of existing
implementations.  They allow for a wide range of code size/efficiency
tradeoffs.

For example, addresses dominate an uncompressed IPv6 header. Level 0
devices only support compressed source addresses, while level 1
devices support all stateless address compression.  In one early
design of this spectrum, Level 0 supported only uncompressed packets.  However, this
raises a problem with ICMP error generation. If a node cannot
decompress the source address of a received packet, it cannot send
ICMP errors. ICMP errors are required for capability discovery.
Stateful compression depends on an out-of-band signal to set up state,
such that nodes only send statefully compressed packets to nodes who
also support it. Therefore Level 0 has the minimal requirement that it
can decompresses stateless source addresses.

The classes in this scale do not precisely reflect the current feature
support of the implementations described in Section~\ref{s:interop}.
For example, Contiki supports UDP port compression (level 5) but does
not support 802.15.4 short addresses (level 2) or tunneled IPv6 (level
5): following this formulation, Contiki only provides level 1 support.
If Contiki supported 16-bit addresses, it would provide level 4
support.

A concrete spectrum such as the one above gives stack designers a
structure and set of guidelines on which features to implement. Based
on our experiences developing a 6LoWPAN stack, we believe that if this
scale existed as part of the initial specification, implementations
would have made an effort to adhere to it. It provides a clear order
with which to incorporate features.

One additional advantage of this spectrum is that it allows for some
future modifications to the P6LoWPAN specification without breaking
interoperability between new and old implementations.  For example,
our scale does not include support for Generic Header
Compression~\cite{RFC7400} because none of the open-source stacks we
analyzed implement it.  Despite this, support for this RFC could
easily be added as a new class on this linear scale (as Class 6), and
all lower class implementations would know they were unable to
interoperate with this new, higher order interoperability class.

This spectrum requires that a node store 3 bits of state for each
neighbor. Given that low-power nodes often store ten or more bytes for
each entry in their link table (link quality estimates, addresses,
etc.), this cost is small.

\subsection{Principle 2: Capability Discovery}

We propose two mechanisms by which P6LoWPAN performs capability
discovery: neighbor discovery (ND) and ICMP\@. Neighbor discovery~\cite{RFC4861} is
analogous to ARP in IPv4: it allows IPv6 devices to discover the link
layer addresses of neighboring addresses as well as local gateways.
Devices use neighbor discovery to proactively discover capability
levels and ICMP to detect when incompatible features are used. Of the
two, only ICMP is required; neighbor discovery simply allows a pair of
nodes to avoid an initial ICMP error if they have different capability
levels.

\vspace{1ex}\noindent {\bf ICMP}: We propose adding a new ICMPv6
message type---P6LoWPAN Class Unsupported---which a device sends in
response to receiving a packet compression modes or 6LoWPAN features
it does not understand. This error encodes the device's capability level.
A node receiving such an error updates its link table entry with the
capability level. In the future, any packets sent to that address use
at most the supported level.

\vspace{1ex}\noindent {\bf Neighbor discovery:} We propose adding a
IPv6 ND option that allows a device to communicate its capability
class during network association. The inclusion of a few extra bits in
ND messages would allow all devices that learn neighbor addresses via
ND to also know how to send packets which that neighbor can receive.
When a node uses ND to resolve an IP address to a link layer address,
it leans the supported capability level as well as the link layer
address.  This option minimizes the energy cost of communicating
capabilities.  It is worth noting that RFC 7400 already employs a
similar method for communicating whether devices implement General
Header Compression: adding such an option is clearly
viable.~\cite{RFC7400}

\subsection{Principle 3: Provide Reasonable Bounds}

Section~\ref{s:interop} discussed two unreasonable bounds which affect
6LoWPAN interoperability: the full-MTU (1280 byte) bound on header
decompression and unbounded recursion when decompressing
tunneled IPv6. 

For P6LoWPAN, we propose that header decompression be bounded to 50 bytes.
This bound allows for significant RAM savings in implementations that
decompress first fragments into the same buffer in which the fragment
was originally held. 50 bytes is a good tradeoff between RAM savings
and how frequently we expect such a bound would force packets to be
sent uncompressed. A 50 byte limit allows for transmission of a packet
containing a maximally compressed IP header, a maximally compressed
UDP header, and still leaves room for some IPv6 extension headers.
Constructing a packets which could require more decompression than
this would require extremely rare and unusual circumstances.
This extremely rare cost, however, buys the
hundreds of bytes of RAM that can be saved, and the increased ease of
creating implementations that will always interoperate.

Second, we propose that headers for tunneled IPv6 should not be
compressed. The primary motivation for this feature was from the RPL
protocol~\cite{RFC6550}, as discussed in Section~\ref{ss:Incomplete}.
However, the fact that RPL must tunnel IPv6 in this way is generally
agreed to be a problem and a wart in its design that should be avoided
when possible~\cite{rpl-authors}. This change allows
implementations to avoid recursive functions to decompress these
headers, and instead use simple if/else statements.

\section{Evaluation}
This section evaluates the costs of applying our three principles to 6LoWPAN.
First, can a reasonable set of capability levels provide a good range of implementation
complexity from which a developer can choose? Second, what is the overhead
of the proposed mechanisms? If the mechanisms themselves increase code
size significantly on their own they are not viable. Third, are
the savings afforded by a linear capability spectrum worth the associated limitations?
Ultimately, we find that the incremental costs of these mechanisms is
small, in the worst case requiring 172-388 bytes of additional code. We also find that
the linear spectrum presents noticeable savings in code size, memory usage, and the size
of capability discovery messages.

\subsection{Implementation}

We implemented the proposed P6LoWPAN on the Contiki-NG 6LoWPAN stack. We selected
Contiki-NG for this purpose because it already has the smallest
6LoWPAN stack of those tested, so represents the case where any overheads the
mechanisms introduce would be the most pronounced. Furthermore,
Contiki-NG offers compile-time options for eliding portions of
its 6LoWPAN implementation, indicating it needs to provide flexible
code size/energy tradeoffs to its users. The changes required
modifying 500 lines of code relative to the head of the 4.2 release of
Contiki-NG. Throughout this section, all code sizes provided are the binary
sizes of Contiki-NG compiled with the Texas Instruments CC2650 as the target.

We modified the Contiki-NG stack so it can be compiled to support any
of the 6 capability levels. We did not add additional 6LoWPAN features
which were absent from the original Contiki-NG 6LoWPAN stack. Our
code size numbers therefore represent a conservative lower bound of the total possible
savings. A compile-time option --- CAPABILITY\_LEVEL --- determines
exactly which features of the 6LoWPAN specification are compiled.

We also added ICMP and ND support for capability discovery. The updated stack responds to
incompatible 6LoWPAN messages with an ICMP error, and communicates its
capability level in Router Solicitation messages using the 6CIO
prefix originally defined in RFC 7400~\cite{RFC7400}. It
stores the capability class of each node in its link table, and
automatically compresses IPv6 packets by the maximum amount supported
by the destination.

Finally, we implemented a second modified 6LoWPAN stack in Contiki-NG,
which does not follow the recommendation of using a linear capability spectrum.
In this modified implementation, each node can select any of the 6LoWPAN features it
chooses. We refer to this implementation as FLEX-6LoWPAN.
For this alternative policy, we isolated 26 features of 6LoWPAN as single
bit flags in a 32 bit bitfield. Thus, FLEX-6LoWPAN stores and communicates
capabilities using 4 byte objects. FLEX-6LOWPAN also supports the added
granularity required to maximally compress outgoing messages intended for a device supporting any specific
combination of features. Once again, we did not add back in any 6LoWPAN features which
the Contiki-NG stack did not originally support. This second implementation required
modifying about 300 additional lines of code from the P6LoWPAN implementation.

The code for both of the aforementioned implementations can be found at \textit{removed for anonymity}.
\subsection{Savings and Costs}

Table~\ref{tab:cap_save} shows the size of the original Contiki-NG 6LoWPAN stack 
compiled at each possible
capability level. Each capability level adds between 0.25 and 1.05 kB
of code, and the spectrum enables implementations to cut the size of
the 6LoWPAN stack by up to 45\%. 

The code size cost of capability discovery, using the P6LoWPAN implementation with
the linear capability spectrum, is shown in
~\ref{tab:cap_cost}.  Capability discovery adds 178-388 bytes, a fraction
of the size which implementations can save by supporting lower
capability levels. The code added for communication varies across
capability levels because the number of code paths for ICMP
error generation and compression changes depending on the supported
level.

\begin{table}[t]
\caption{Code size of different capabilities levels in Contiki-NG. Each capability adds
between 0.3 and 1.1 kB, with the spectrum spanning a nearly 100\% increase in code size
from the lowest capability.}\label{tab:cap_save}
\begin{tabular}{lrr}
\toprule
%\bf{Capability}                                        & \multicolumn{3}{c}{ \bf{6LoWPAN Code Size  (kB)} }   &                                   \\
\hline
{\bf{Capability}}	       & {\bf{Code Size (kB)}} & {\bf{Increase (kB)}} \\% & \bf{Elements Missing from Contiki-NG} \\
\hline
Level 0   & 3.4 & -    \\ %&  \\
Level 1   & 4.4 & 1.1  \\ %&  \\
Level 2   & 5.2 & 0.7  \\ %&  Stateful Multicast Compression \\
Level 3   & 5.4 & 0.3  \\ %&    \\
Level 4   & 5.9 & 0.4  \\ %&  Tunneled IPv6, UDP Checksum Elision \\
Level 5   & 6.3 & 0.5  \\ %&  Broadcast Header, Mesh Header, IPv6 Mobility Header, RFC 6775  \\
\bottomrule 
\end{tabular}
\end{table}

\begin{table}[t]
\caption{Cost of implementing capability discovery in Contiki-NG. On average,
the cost of discovery is less than 5\% of the total 6LoWPAN size, and capability
levels offset the overhead of capability discovery by a factor of 10.}\label{tab:cap_cost}
\begin{tabular}{lrrr}
\toprule
\bf{Capability}                                        & \multicolumn{2}{c}{ \bf{6LoWPAN Code Size  (kB)} }      \\
	       & \bf{Base}   & \bf{w/Discovery} & \bf{Increase} \\
\hline
Level 0  & 3.2      & 3.4  & 188 bytes  \\ %2.870 if src addr decomp is part of comm code
Level 1  & 4.2      & 4.4  & 260 bytes   \\
Level 2  & 4.8      & 5.2  & 388 bytes   \\
Level 3  & 5.1      & 5.4  & 340 bytes   \\
Level 4  & 5.6      & 5.9  & 296 bytes   \\
Level 5  & 6.2      & 6.3  & 172 bytes    \\
\bottomrule 
\end{tabular}
\end{table}

Finally, we evaluated the importance of using a linear capability spectrum by comparing our P6LoWPAN
implementation with our bitfield-based FLEX-6LoPWAN implementation. While FLEX-6LoWPAN allows for
a completely tailored 6LoWPAN on each node, it has several immediately obvious drawbacks. This approach 
requires communicating 32 bits of state to convey
or store a nodes capability, instead of 3 bits as in the case of a linear spectrum. This represents a relatively
small increase in RAM usage by 6LoWPAN, but a significant increase in communication cost - 4 bytes is 
a substantial added length, and makes capability communication using short ND options like the 6CIO option
impossible, forcing the use of some longer, currently unspecified ND option. It also
makes the process of determining the maximum allowable compression between two nodes significantly
more complex, as demonstrated in Table~\ref{tab:linear vs. field}. One important takeaway from this table
is that opting for a less restrictive set of allowable capabilities mitigates much of the savings
provided by implementing these capabilities. For example, a FLEX-6LoWPAN device with the equivalent 
of level 4 capabilities requires more code space than a level 5 P6LoWPAN device -- the linear capability spectrum makes a difference.
Once again, the code size addition for FLEX-6LoWPAN represents a conservative lower bound, as
we did not need to add checks for handling the compression features of 6LoWPAN that Contiki-NG does
not support.

\begin{table}[t]
\caption{Resource requirements for a 6LoWPAN stack in Contiki-NG using a linear capability spectrum vs. using a capability bitfield
that allows nodes to pick and choose arbitrary features from the original specification. These numbers are for a node configured with capabilities equivalent to a level 4 device. The capability field approach adds over 0.5 kB of code size, increases ND router solicitations by 4 bytes, and increases RAM used for capability storage.}\label{tab:linear vs. field}
\begin{tabular}{lrr}
\toprule
\hline
   --	 & {\bf{Linear Spectrum}} & {\bf{Arbitrary Bitfield}} \\
\hline
6LoWPAN Code Size   & 5.9 kB &  6.5 kB   \\ % Precise difference is 570 Bytes
Size of ND Option   & 4 Bytes & 8 Bytes  \\
RAM per neighbour  & 19 Bytes & 22 Bytes  \\ % (16 bytes is the neighbour IP address)
\bottomrule 
\end{tabular}
\end{table}

\section{Discussion and Conclusion}
A new generation of low-power devices face a connectivity dilemma: Internet protocols
are not designed for energy efficiency, but compression and other energy saving
adaptations takes up precious code space. Device deployments
specialized for single-vendor local networks make trade-offs specific to their
application requirements. As a result, IP communication between IP enabled devices
fails. 

These trade-offs are fundamental to low-power networked devices. Popular and
heavily used 6LoWPAN stacks today implement different subsets of the protocol.
Some of them even include compile-time configuration options to elide additional required
features in order to further conserve code space and fit the stack into a particular
application. Arguing that every device should always implement everything will
relegate many devices and applications to the brittle, proprietary protocols they
have been historically so problematic.

Part of the challenge is that some traditional protocol design principles do not apply
well to the low-power setting. While senders are given flexibility in how to format
a message, receivers are expected to be able to process any particular sender's decision.
This asymmetry in expectations forces implementations to cover every case which is 
more complex than is feasible for many low-power devices.

We propose three new design principles for low-power Internet protocols. These principles
explicitly acknowledge the unique code space/energy tradeoffs of low-power devices by
allowing individual devices to choose their own point in the tradeoff space while
simultaneously allowing devices with different tradeoffs to find common ground:

\begin{enumerate}
    \item \textbf{Capability spectrum}: acknowledge that low power devices will have varying abilities to implement complex protocols and define a linear spectrum
    of levels such that the costs of complex techniques are only borne by more resourceful devices. 
    \item \textbf{Capability discovery}: include a mechanism for determining which level each endpoint supports. Ensure that such a mechanism is primarily reactive, but
    that discovery requires no more than one failure.
    \item \textbf{Explicit and finite bounds}: static bounds should exist to establish the maximum resources a packet may consume. Bounds should be conservative, finite, and, above all else, pervasive.
\end{enumerate}

While traditional network principles such as layering and end-to-end still have their place 
in low power networks, these three principles are also paramount. Without them, traditional
network principles can fail to provide devices with the effortless connectivity that is the
hallmark of the Internet's success. We evaluate these principles through an example
application to the 6LoWPAN protocol
suite, where we find they prevent communication failures with at most a few hundred bytes of
overhead.

Looking forward, considering the tension between energy efficiency and code size 
is critical for protocol designers in this ecosystem of diverse hardware capabilities
and application tradeoffs. 6LoWPAN is not
the only low power Internet protocol --- the low power space uses its own routing protocols, 
address discovery protocols, and application layer protocols ~\cite{RFC6550, RFC7252}. Additional
protocols will follow as the space matures. Many of these protocols will be
initially developed outside the IETF --- Jonathan Hui was a graduate student
when he presented the first complete IPv6-based network architecture for sensor nets~\cite{hui}. We
have presented a roadmap for how these principles 
can reframe the discussion of how to connect the next hundred billion devices
to the Internet.

\balance

\bibliographystyle{ACM-Reference-Format}
\bibliography{main}

%\appendix

\end{document}